\documentclass[]{el-author}

\usepackage{bm}
\usepackage{cancel}

\DeclareMathOperator{\Tr}{trace}

\newcommand{\singlePlotsSize}{55mm}
\newcommand{\doublePlotsSize}{61mm}

\newcommand{\state}{\theta}
\newcommand{\bfState}{\bm{\state}}
\newcommand{\lenState}{K}

\newcommand{\obs}{y}
\newcommand{\bfObs}{\textbf{y}}
\newcommand{\lenObs}{N}

\newcommand{\nSamples}{M}
\newcommand{\samplesSet}{\Theta^\nSamples}
\newcommand{\nClip}{\nSamples_T}

\newcommand{\nObs}{N}

\newcommand{\mixCoeff}{\rho}

\newcommand{\iMCtrial}{r}
\newcommand{\iObsTrial}{l}

\newcommand{\nMCtrials}{\MakeUppercase{\iMCtrial}}
\newcommand{\nObstrials}{\MakeUppercase{\iObsTrial}}

\newcommand{\tiwAlg}{NIS}

\newcommand*{\expectSymbol}{{\mathbb E}}
\newcommand*{\expect}[2]{\expectSymbol_{#2}\left[#1\right]}

\newcommand{\sample}[1]{\bfState^{(#1)}}
\newcommand{\weight}[1]{w^{(#1)}}
\newcommand{\uweight}[1]{w^{(#1)*}}
\newcommand{\tweight}[1]{\bar w^{(#1)}}
\newcommand{\utweight}[1]{\bar w^{(#1)*}}
\newcommand{\transf}[1]{\varphi_{\samplesSet}\left(#1\right)}
\newcommand{\gaussian}[3]{{\mathcal N}\left(#1|#2,#3\right)}
\newcommand*{\est}[1]{\hat{#1}}
\newcommand*{\modulus}[1]{\left\|  #1 \right\|}
\newcommand*{\mean}[1]{\bar{#1}}
\newcommand*{\trace}[1]{\Tr\left\lbrace #1\right\rbrace}

\begin{document}

\title{Importance sampling with transformed weights}

\author{Manuel A. V\'azquez and Joaqu\'in M\'iguez}

\abstract{

The importance sampling (IS) method lies at the core of many Monte Carlo-based techniques. 
IS allows the approximation of a target probability distribution by drawing samples from a proposal (or \textit{importance}) distribution, different from the target, and computing importance weights (IWs) that account for the discrepancy between these two distributions. The main drawback of IS schemes is the degeneracy of the IWs, which significantly reduces the efficiency of the method. It has been recently proposed to use transformed IWs (TIWs) to alleviate the degeneracy problem in the context of Population Monte Carlo, which is an iterative version of IS.
However, the effectiveness of this technique for standard IS is yet to be investigated. In this letter we numerically assess the performance of IS when using TIWs, and show that the method can attain robustness to weight degeneracy thanks to a bias/variance trade-off.
}

\maketitle

\section{Introduction}

One classical application of Monte Carlo (MC) methods is  the approximation of a distribution of interest (often referred to as \textit{target} distribution) by means of random samples. In many practical situations it is not possible or convenient to draw samples directly from the target. In such a case, it is common to rely on the importance sampling (IS) principle \cite{Robert04}. It consists in drawing samples from a \textit{proposal} distribution, which are then assigned importance weights (IWs) to compensate for the mismatch between the target and the proposal.

A critical drawback of the IS methodology is the degeneracy of the IWs. This happens when only a few samples have non-negligible IWs. Since samples with weights close to zero are irrelevant when building a Monte Carlo approximation, weight degeneracy reduces the efficiency of the method. This problem aggravates when performing inference in high-dimensional systems \cite{bengtsson2008curse}.

The term Population Monte Carlo (PMC) \cite{Cappe2004} refers to a class of iterative IS algorithms in which samples drawn from a proposal distribution are used to obtain a refined proposal that can be sampled again. In the context of PMC, the authors of \cite{koblents2015population} propose to apply a non-linear transformation to the IWs that reduces their variability, and hence alleviates the degeneracy problem. Although a theoretical analysis of the asymptotic convergence of the method is provided in \cite{koblents2015population}, to date there is no published numerical assessment of the performance of the importance samplers with transformed IWs (TIWs) in a non-iterative setting and with finite sample size. Moreover, the impact of certain parameters that are relevant to the performance of the method has not been investigated.
In this work we tackle these open issues.

\section{Standard Monte Carlo}

Let $\bfState$ be an unknown $\lenState \times 1$ random vector with known prior density $p(\bfState)$. Our goal is the Bayesian estimation of $\bfState$ given an $\lenObs \times 1$ vector of observations, $\bfObs$, that relates to the former through a likelihood function, $p(\bfObs | \bfState)$. Specifically, we aim at approximating the posterior probability density function (pdf) of $\bfState$, i.e.,
\begin{equation}
	p(\bfState | \bfObs)
	\propto
	p(\bfObs | \bfState)
	p(\bfState)
	\label{eq:posterior}
\end{equation}
using a collection of $\nSamples$ random samples, $\samplesSet = \{ \sample{i} \}_{i=1}^\nSamples$, in the space of $\bfState$. From the latter, it is easy to approximate any expectation of the form $\expect{f(\bfState)}{p(\bfState | \bfObs)} = \int f(\bfState)p(\bfState | \bfObs)d\bfState$, where $f: \mathbb{R}^{\lenState} \to \mathbb{R}$ is some real integrable function of $\bfState$. For instance, the posterior mean of $\bfState$ can be approximated as $\expect{f(\bfState)}{p(\bfState | \bfObs)} \approx \frac{1}{\nSamples} \sum_{i=1}^{\nSamples} \sample{i}$.

\section{Importance sampling}

Let us denote by $\pi(\bfState)$ the pdf of a distribution of interest, usually referred to as the \textit{target} distribution. It is often impractical to sample $\pi(\bfState)$ directly, so we are going to draw samples from a \textit{proposal} distribution and weight them appropriately according to the principle of IS \cite{Robert04}. If $q(\bfState)$ denotes the proposal pdf, then the idea is to draw samples, 
\begin{equation}
	\sample{i} \sim q(\bfState),\, i=1,\cdots,\nSamples
	\label{eq:sampling}
\end{equation}
and assign each one a normalized importance weight, $\weight{i}$, 
computed as
\begin{equation}
	\begin{array}{ccc}
		\uweight{i}
		\propto
		\frac{\pi(\sample{i})}{q(\sample{i})}
		,& 
		\weight{i}
		=
		\frac{
			\uweight{i}
		}{
			\sum_{i=1}^{\nSamples} \uweight{i}
		}
		,& 
		i=1,\cdots,\nSamples
	\end{array}
	.
	\label{eq:weight_computation}
\end{equation}
Equations \eqref{eq:sampling} and \eqref{eq:weight_computation} together constitute the standard IS algorithm.

From the sample set $\samplesSet=\{\sample{i}\}_{i=1}^\nSamples$ and their associated weights, one can build a discrete random measure,
\begin{equation}
	\pi^\nSamples(d\bfState)
	=
	\sum_{i=1}^{\nSamples}
	\weight{i}
	\delta_{\sample{i}}(d\bfState)
	,
	\label{eq:pdf_approximation}
\end{equation}
where $\delta_{\sample{i}}$ is the unit delta measure located at $\bfState = \sample{i}$, that allows to approximate the expectation of any integrable function $f$ with respect to $\pi(\bfState)$ as
\begin{equation}
	\expect{f(\bfState)}{\pi(\bfState)}
	\approx
	\sum_{i=1}^{\nSamples}
	\weight{i}
	f(\sample{i})
	.
	\label{eq:approximation}
\end{equation}

The efficiency of any IS algorithm (roughly given by the number of samples that are required to attain a certain level of performance) depends to a great extent on the choice of the proposal function, $q(\bfState)$.
However, the asymptotic convergence of the above approximation when $\nSamples \to \infty$ is guaranteed as long as 
$
	0
	<
	\frac{\pi(\bfState)}{q(\bfState)}
	<
	\infty
$
for every $\bfState$ \cite{Robert04}.

\section{Weight degeneracy and transformation of the weights}

The degeneracy of the IWs refers to the situation in which only a few samples hold significant IWs, while the vast majority have weights close to zero. This clearly reduces the efficiency of the method since samples with negligible IWs barely contribute to the approximations in \eqref{eq:pdf_approximation} or \eqref{eq:approximation}.

Following \cite{koblents2015population}, we propose to alleviate this problem by applying a non-linear transformation over the unnormalized IWs, $\uweight{i}$, aimed at decreasing their variance. To be specific, the (unnormalized) transformed IWs (TIWs) are obtained as $\utweight{i} = \transf{\uweight{i}}, i=1,\cdots,\nSamples$, where $\transf{\cdot}$ is a real positive function that depends on the whole sample set, $\samplesSet$, and their weights.

Different choices for the non-linear transformation, $\transf{\cdot}$, are possible. The asymptotic convergence results in \cite{koblents2015population} are specifically referred to the "clipping" transformation, which is also the one we consider here. It consists in setting the $\nClip < \nSamples$ highest IWs to a common value. More formally, let us consider a permutation, $i_1,\cdots,i_\nSamples$ of the indices in $\{1,\cdots,\nSamples\}$ such that $\uweight{i_1} \ge \cdots \ge \uweight{i_{\nClip}} \ge \cdots \ge \uweight{i_\nSamples}$. Then, the unnormalized TIWs, $\utweight{i}$, are computed from the original IWs, $\uweight{i}$, as
\begin{equation}
	\utweight{i}
	=
	\transf{\uweight{i}}
	=
	\min
	\left(
		\uweight{i}
		,
		\uweight{i_{\nClip}}
	\right)
	,
	\label{eq:transformed_weight_computation}
\end{equation}
and their normalized counterparts are
\begin{equation}
	\tweight{i}
	=
	\frac{
		\utweight{i}
	}{
		\sum_{i=1}^{\nSamples} \utweight{i}
	}
	,
	i=1,\cdots,\nSamples
	.
	\label{eq:normalized_transformed_weight_computation}
\end{equation}
The IS algorithm with TIWs is described by equations \eqref{eq:sampling}, \eqref{eq:weight_computation}, \eqref{eq:transformed_weight_computation} and \eqref{eq:normalized_transformed_weight_computation}. We refer to it as nonlinear IS (\tiwAlg{}).

\section{Simulations}

We numerically asses the convergence of the \tiwAlg{} algorithm when the sample size is finite and, most importantly, when using a non-iterative scheme (unlike in \cite{koblents2015population}).
We apply the algorithm to estimate the location of the modes of a Gaussian mixture model (GMM) given a set of conditionally independent and identically distributed (i.i.d.) observations. In particular, we consider the GMM
\begin{equation}
	p(\obs|\bfState)
	=
	\mixCoeff_1
	\gaussian{y}{\state_1}{\sigma^2} +
	\mixCoeff_2
		\gaussian{y}{\state_2}{\sigma^2} +
	(1-\mixCoeff_1 - \mixCoeff_2)
	\gaussian{y}{\state_3}{\sigma^2}
	,
	\label{eq:GMM}
\end{equation}
where $\bfState = \left[\state_1, \state_2, \state_3\right]^\top = \left[0, 2, 4\right]^\top$ is a vector encompassing the unknown means of the mixture components. The remaining parameters of the model are assumed to be known, and set to $\mixCoeff_1=0.2$, $\mixCoeff_2=0.3$ and $\sigma^2 = 1$.

We assume independent Gaussian prior distributions for the unknown means, namely
\begin{equation}
	p(\bfState)
	=
	\gaussian{\state_1}{1}{10\sigma^2}
	\gaussian{\state_2}{1}{10\sigma^2}
	\gaussian{\state_2}{1}{10\sigma^2}
	.
\end{equation}

Given a collection of $\nObs$ observations, $\bfObs = \left\lbrace \obs_1,\obs_2,\cdots,\obs_\nObs\right\rbrace$,  drawn from the GMM in \eqref{eq:GMM}, we aim at approximating the posterior $\pi(\bfState|\bfObs)$. In order to do so, we draw samples from a proposal function, which is selected to match the prior, i.e.,
\begin{equation}
	\sample{i}
	\sim
	q(\bfState)
	=
	p(\bfState)
	,
	i=1,\cdots,\nSamples
	.
\end{equation}
Then, at the sight of equations \eqref{eq:weight_computation} and \eqref{eq:posterior}, the unnormalized IWs are computed as
\begin{equation}
	\uweight{i}
	\propto
	\frac{\pi(\bfState)}{q(\bfState)}
	\propto
	p(\bfObs | \bfState)
	=
	\prod_{i=1}^{\nObs}
	p(\obs_i|\bfState)
	,
\end{equation}
where we have used the fact that the observations in $\bfObs$ are conditionally independent given the state. TIWs can then be computed as indicated in \eqref{eq:transformed_weight_computation} and \eqref{eq:normalized_transformed_weight_computation}.

In this setting we compare the performance of the standard IS and the novel NIS schemes. Every result in this section is averaged over a certain number of independent simulation runs, $\nMCtrials$, of the appropriate algorithm for a fixed set of observations $\bfObs$. We refer to each of these runs as an MC realization, and we set $\nMCtrials=1,000$. Additionally, $\nObstrials$ independent realizations of the observations, $\bfObs_1,\bfObs_2,\cdots,\bfObs_{\nObstrials}$, are considered. Overall, each algorithm is run $\nObstrials\nMCtrials$ times.

\begin{figure}[htpb]
	\centering{
		\includegraphics[width=\singlePlotsSize]{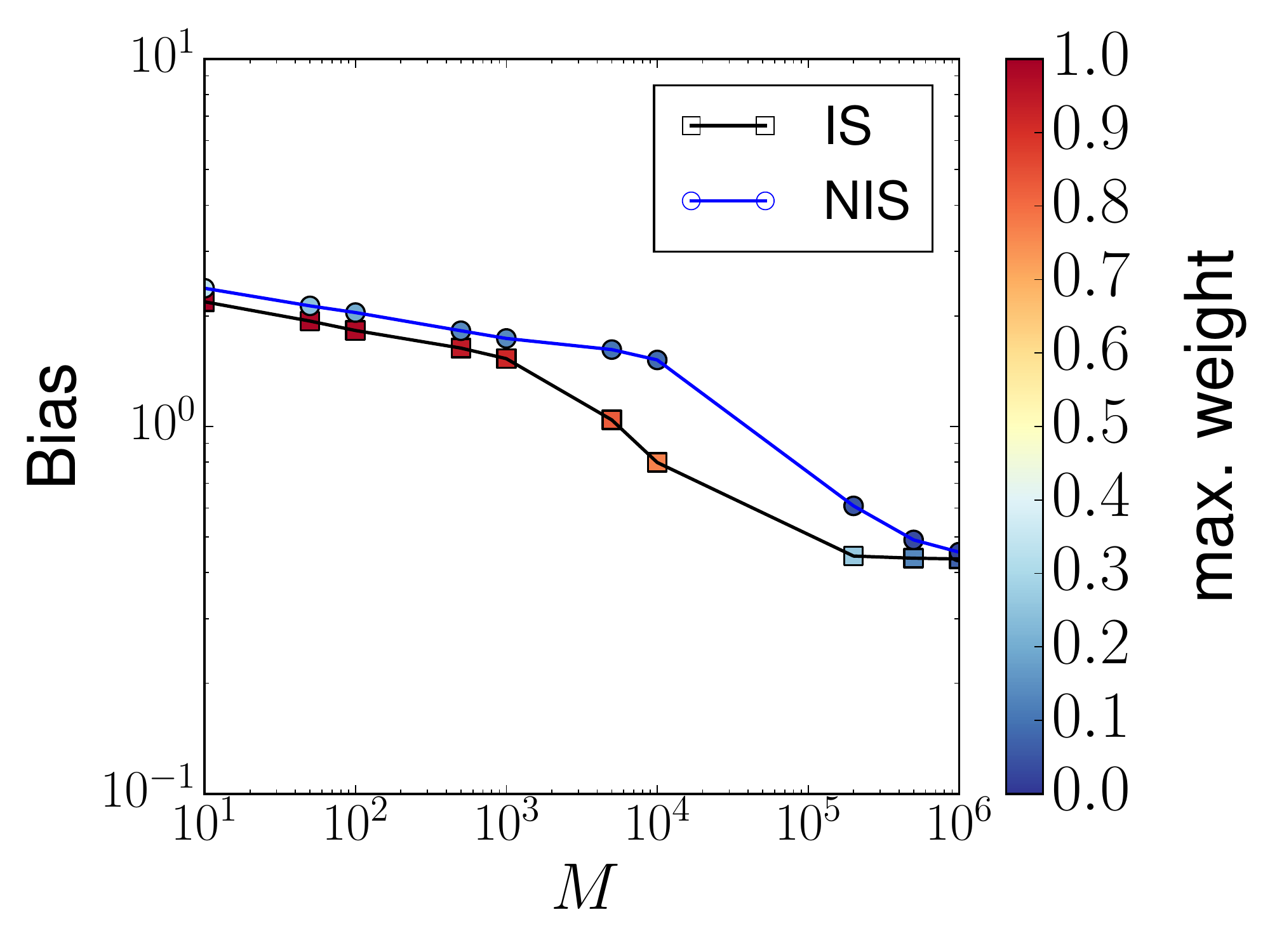}
	}
	\caption{Bias averaged over $\nObstrials=75$ observations realizations.}
	\label{fig:bias}
\end{figure}

\begin{figure}[htpb]
	\centering{
		\includegraphics[width=\singlePlotsSize]{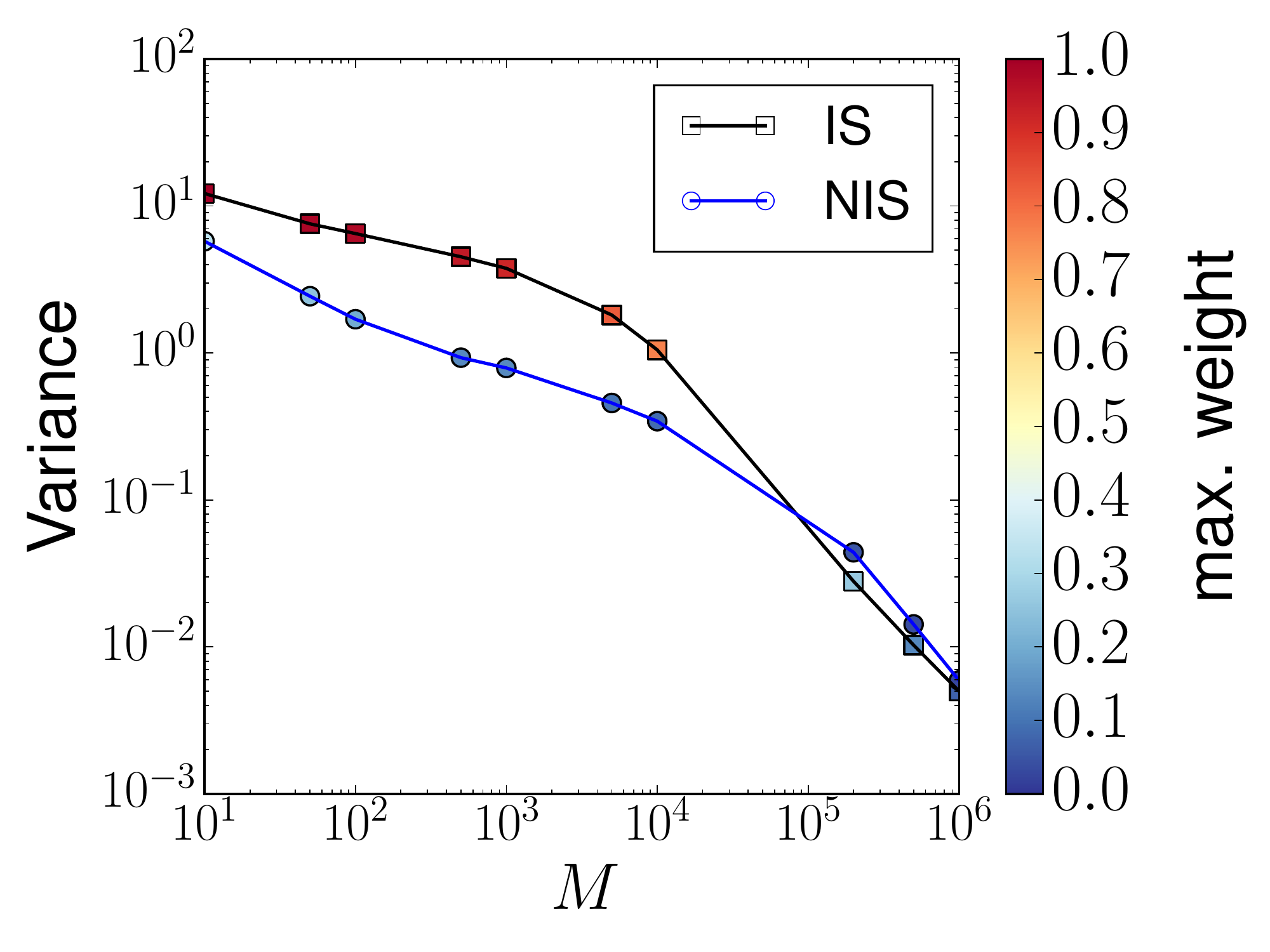}
	}
	\caption{Variance averaged over $\nObstrials=75$ observations realizations.}
	\label{fig:variance}
\end{figure}

Figure \ref{fig:bias} shows the performance of the standard IS and \tiwAlg{} algorithms
in terms of the bias of the posterior-mean estimators, 
$
	\est{\bfState}_{NIS}
	=
	\sum_{i=1}^{\nSamples}
	\tweight{i}
	\sample{i}
$
and
$
	\est{\bfState}_{IS}
	=
	\sum_{i=1}^{\nSamples}
	\weight{i}
	\sample{i},
$
as the number of samples, $\nSamples$, grows. For every value of $\nSamples$, the number of samples whose weights are clipped is $\nClip = \log(\nSamples)$. On the other hand, the number of observations is in every case $\nObs=1,000$. The bias is computed as
\[
	\text{Bias}
	=
	\frac{1}{\nObstrials}
	\sum_{\iObsTrial=1}^{\nObstrials}
	\modulus{
		\frac{1}{\nMCtrials}
		\sum_{\iMCtrial=1}^{\nMCtrials}
		\est{\bfState}_{\iObsTrial,\iMCtrial} -
		\bfState
	}
	,
\]
where $\est{\bfState}_{\iObsTrial,\iMCtrial}$ is the posterior-mean estimate of $\bfState$ computed for the $\iObsTrial$-th realization of the observations, $\bfObs_\iObsTrial$, during the $\iMCtrial$-th MC realization for either the \tiwAlg{} or IS algorithms.
Additionally, the figure also provides information about the degeneracy of the weights. The color of every marker indicates, according to the color bar on the right, the maximum weight among all the samples.
It can be seen that the bias attained by the standard IS estimator is always below the bias of the \tiwAlg{} estimator. This is because the transformation of the weights applied to obtain the TIWs introduces a distortion in the approximation of the probability measure $\pi(\bfState)d\bfState$.
However, it is clear from the figure that both algorithms converge to the same estimate when the number of samples is large as predicted by the asymptotic analysis in \cite{koblents2015population}. 
Also notice that in the standard IS algorithm, the maximum weight is close to $1$ whenever the number of samples is below $\nSamples=10,000$. 
Hence, most of the probability mass is concentrated in a single sample, and the remaining samples are, for all practical purposes, irrelevant. 
This is specifically avoided in the \tiwAlg{} algorithm, where the maximum weight is below $0.2$ for most values of $\nSamples$.
Avoiding a single sample garnering most of the probability mass is important in connection to the variance of the resulting estimators. This is illustrated in Figure \ref{fig:variance} that shows the estimator variance, computed as
$
	\text{Variance}
	=
	\frac{1}{\nObstrials}
	\sum_{\iObsTrial=1}^{\nObstrials}
	\trace{
		\frac{1}{\nMCtrials}
		\sum_{\iMCtrial=1}^{\nMCtrials}
		\left(
			\est{\bfState}_{\iObsTrial,\iMCtrial} -
			\mean{\est{\bfState}}_{\iObsTrial}
		\right)
		\left(
			\est{\bfState}_{\iObsTrial,\iMCtrial} -
			\mean{\est{\bfState}}_{\iObsTrial}
		\right)^\top
	}
$
with
$
	\mean{\est{\bfState}}_{\iObsTrial}
	=
	\frac{1}{\nMCtrials}
	\sum_{\iMCtrial'=1}^{\nMCtrials}
	\est{\bfState}_{\iObsTrial,\iMCtrial'}
	.
$
The figure shows that, for low-to-medium sample sizes, the variance of the estimates computed using the \tiwAlg{} method is considerably lower than the variance of the standard IS estimates. 

A common metric for comparing the performance of different algorithms is the mean square error (MSE). Our simulations indicate that, e.g., for $\nSamples=1,000$ the (average) MSE attained by the standard IS algorithm is $6.21$ while that achieved by the \tiwAlg{} scheme is $3.82$.

\begin{figure}[htpb]
	\centering{
		\includegraphics[width=\doublePlotsSize]{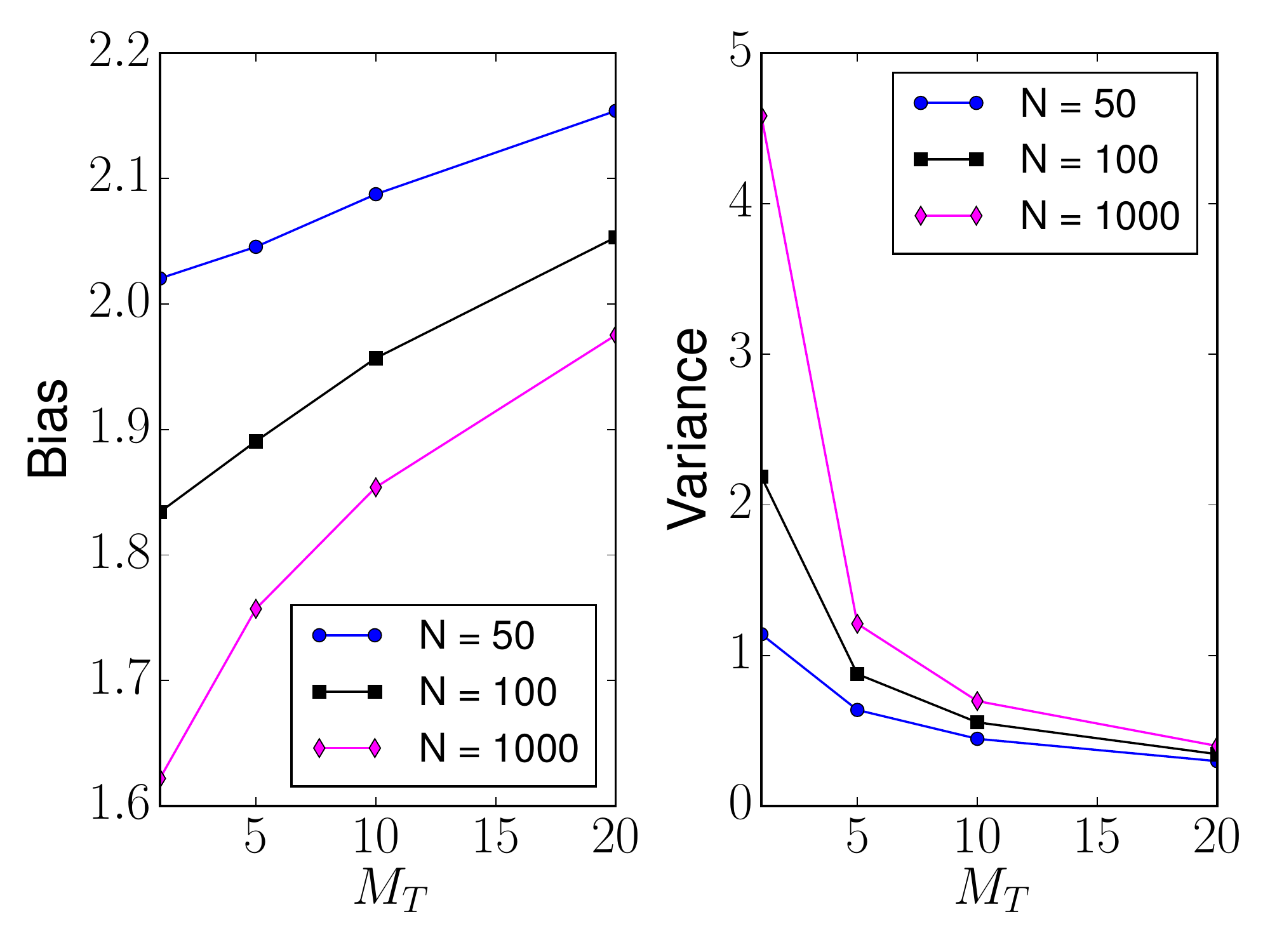}
	}
	\caption{Bias (Left) and variance (Right) obtained by the \tiwAlg{} algorithm for $\nSamples=500$. The results are averaged over $\nObstrials=200$ observations realizations.}
	\label{fig:biasVariance}
\end{figure}

In the last experiment, we explore the impact on the performance of the \tiwAlg{} scheme of both the number of observations, N, and the number of clipped weights, $\nClip$. Notice that when $\nClip=1$, the resulting algorithm is the standard importance sampler. Figure \ref{fig:biasVariance} shows that, for any number of observations $\nObs$, the bias (left) increases along $\nClip$, whereas the variance (right) decreases. However, there is an elbow in the curves for the variance, which suggests that increasing the value of $\nClip$ above $10$ does not yield any benefits in terms of variance while the bias keeps increasing linearly. Another remarkable result is that, as $\nObs$, grows, the bias decreases and the variance increases.
This is due to the target pdf concentrating in an ever smaller region as the number of observations, $\nObs$, grows, which, in turn, makes sampling more difficult. In such a case, the benefits stemming from using \tiwAlg{} are more obvious. This can be seen by comparing, e.g., the variance when $\nClip=1$ (plain IS) and $\nClip=5$ for $\nObs=1,000$ observations.

\section{Conclusion}
We have investigated the benefits of applying a nonlinear transformation to the IS weights in order to alleviate the well-known degeneracy problem. 
Our computer simulations show that, while both the IS and \tiwAlg{} schemes converge to the same approximations when the number of samples is large enough, the estimators computed via the \tiwAlg{} method attain an advantageous variance/bias trade-off that often results in a better practical performance.

\vskip3pt
\ack{This work was supported by {\em Ministerio de Econom\'{\i}a y Competitividad} of Spain (projects TEC2012-38883-C02-01  and TEC2015-69868-C2-1-R)  and the Office of Naval Research Global (award no. N62909- 15-1-2011).}

\vskip2pt

\noindent M. A. V\'azquez and J. M\'iguez (\textit{Departamento de Teor\'{\i}a de la Se\~nal y Comunicaciones, Universidad Carlos III de Madrid, Spain})

\vskip2pt

\noindent E-mail: \{mvazquez,jmiguez\}@tsc.uc3m.es

\bibliographystyle{elsarticle-num}
\bibliography{biblio}

\end{document}